\documentclass[english,prb,twocolumn,aps,amsmath,amssymb,notitlepage,superscriptaddress,showpacs]{revtex4-1}
\usepackage{graphicx}
\usepackage{hyperref}
\usepackage{amssymb}
\usepackage{amsmath}
\usepackage{setspace}
\usepackage{babel}
\usepackage{color}
\newcommand{\beq}{\begin{equation}}
\newcommand{\enq}{\end{equation}}
\def\bf#1{\pmb{#1}}
\def\l{\left}
\def\r{\right}

\def\ket#1{\left |#1\right\rangle}  
\def\bra#1{\left\langle #1\right|}

\begin{document}
%
\title{$Ab$-$initio$ electronic structure, optical and magneto-optical properties \\ of $MnGaAs$ digital ferromagnetic heterostructures}

\newcommand{\ism} {CNR-ISM, U.o.s. di Montelibretti, via Salaria Km 29.3, I-00016 Montelibretti, Italy, European Union}
\newcommand{\mdm} {CNR-IMM, U.o.s. Agrate Brianza, via Olivetti 2, I-20864,  Agrate Brianza, Italy, European Union}
\newcommand{\etsf} {European Theoretical Spectroscopy Facilities (ETSF), Italy}
\newcommand{\unimi} {Dipartimento di Fisica, Univesit\`{a} degli studi di Milano, via Celoria 16, IT 20133 Milano, Italy, European Union}

\author{Patrizia Rosa}
\affiliation{\unimi}
\affiliation{\mdm}
\author{Davide Sangalli}
\affiliation{\mdm}
\affiliation{\ism}
\affiliation{\etsf}
\author{Giovanni Onida}
\affiliation{\unimi}
\affiliation{\etsf}
\author{Alberto Debernardi\footnote{e-mail:alberto.debernardi@mdm.imm.cnr.it}}
\affiliation{\mdm}
%
\begin{abstract}
We report on a theoretical study of the electronic, optical and magneto--optical properties of digital ferromagnetic hetero--structures based on $Mn$ $\delta$--doped $GaAs$. We consider different structures corresponding to $Mn$ contents within the range 12-50\% and we study how the system changes as a function of the doping concentration. Our first--principles approach includes the spin--orbit interaction in a fully relativistic pseudopotential scheme and the local fields effect in the description of the optical absorption. We show that $Mn$ $\delta$--doped $GaAs$ shares many properties with the uniformly doped $Ga_{1-x}Mn_xAs$ system, i.e. half--metallicity, similar absorption spectra, and moderate Kerr rotation angles in the visible spectral region.
\end{abstract}
\maketitle
\section{Introduction}
 
Digital Ferromagnetic Heterostructures (DFH) have been recently proposed as a way to overcome some of the drawbacks
of random--doped $Ga_{1-x}Mn_xAs$ Diluted Magnetic Semiconductors (DMS), without loosing the peculiar properties of the latter connected with the possibility to exploit spin--polarization effects within a semiconducting (or half--metallic) medium compatible with the present technology of semiconductor electronic devices  \cite{ReviewDMS-Sato1}.
$Mn$--based DFH in III-V semiconductors are super--lattice heterostructures where (fully or partially filled) mono-layers (ML)  of $Mn$ --which is substitutional to cation group III elements-- are spaced by a fixed number of layers of the III-V compound. 
On one hand, these $\delta$-doped structures allow a concentration of $Mn$ higher than in random-doped ($Mn$-)III-V semiconductors, and thus, 
due to the different valence between the group III atom and the substitutional $Mn$, 
can produce in principle a larger concentration of holes, enhancing the Curie temperature ($T_C$) of the system.
On the other hand, DMS $T_C$ can increase considerably also because it depends on the spatial distribution of magnetic ions, which can change substantially in low dimensional systems: Nazmul and co-workers~\cite{Nazmul1} showed that in a $p$-type selectively doped III-V hetero-structured composed of $Mn$ $\delta$-doped $GaAs$/$p$-$AlGaAs$ a remarkably high $T_C$ of about 250 K can be reached; 
in Ref.~[\onlinecite{Chen1}] Chen {\it et al.} report $T_C$ well above room temperature in $GaSb/Mn$ digital alloys.   

Possible applications of DMS range from spintronics (i.e., the idea to use the spin degree of freedom in order to transmit, write, or store digital data) \cite{Wolf1,Behin_Aein1} to the use of spin--polarized currents for the study of surface magnetism and magnetic anisotropies in materials. 
In this respect, half metals, which are conductive in one spin channel and semiconducting in the other, are ideal candidates to build spin--injectors. Many DMS have indeed an half-metallic behavior~\cite{ReviewDMS-Sato1}. 
Moreover, several DMS also display intriguing magneto--optical coupling effects: an example is the Magneto Optical Kerr Effect~\cite{kojima} (MOKE), i. e. the polarization rotation of the light reflected by the surface of a magnetic material, which is at the basis of the operation of magneto--optical data storage media

Moreover, the MOKE can be used in the study of electronic structure in magnetic materials. 

Previous studies of $GaAs$ based DFH include the experimental realization, by means of low--temperature Molecular Beam Epitaxy (MBE), of $GaAs$/$Ga_{1-x}AsMn_x$, separated by $GaAs$ spacers of variable thicknesses~\cite{AuthorGaAsMn_Exp1}, and measurements of their optical conductivity spectra in the infrared region~\cite{AuthorGaAsMn_Exp2}.

On the theoretical point of view the optical and magneto--optical properties of $Mn$--doped $GaAs$ have been studied by 
	means of the linearized Full--Potential methods~\cite{Picozzi1,Stroppa1} (FLAPW approach) or
by the Korringa--Kohn--Rostoker method~\cite{Ogura1}, but only in the case of an 
	uniform $Ga_{1-x}Mn_{x}As$ doping.
	Ab-initio theoretical predictions for the magneto--optical properties of DFH are hence lacking.

	A possible advantage of DFH $\delta$-doped structures with respect to uniformly doped $Ga_{1-x}Mn_{x}As$ DMS derives from 
	the possible amplification of MOKE effects by exciton confinement. However, also local field effects (corresponding indeed to the electron-hole exchange in the excitonic Hamiltonian~\cite{Onida1} ) may play an important role in such a strongly anisotropic system.
	However, the intrinsic complexity of DFH, with unit cells containing many atoms, makes the direct application of full-potential methods such as those of Refs.~[\onlinecite{Stroppa1,Picozzi1}] computationally prohibitive.  

	On the other hand, more efficient ab--initio methods based on pseudo--potentials (PP) and plane waves have not been systematically used for the study of MOKE spectra, because of the common wisdom that an all--electron approach was required in order to describe the wave--function within the core region (where the spin-orbit (SO) interaction is stronger). Instead,
	some of us have recently demonstrated that, provided that the spin-orbit interaction is fully taken into account in the construction and use of ``fully relativistic'' PP, the plane--waves based approach provides results of the same quality as those of all--electron calculations~\cite{Sangalli1}.

	Indeed the MOKE is due to a different dielectric response for right ($\epsilon_+$) and left  ($\epsilon_-$) circularly polarized light, which stems from the difference between the wave--functions with $+L_z$ character and $-L_z$ character. In a magnetic material this difference is due to the SO coupling in the Hamiltonian, $H=\xi \mathbf{L\cdot S}$, which transfer the spontaneous symmetry breaking between the $+S_z$ and the $-S_z$ spin component to the spatial part of the wave--function. Thus the MOKE cannot be captured by a perturbative description of the SO effect, nor by the use of scalar relativistic PP, which only include the effect of the SO in the energy levels without affecting the wave--function.

	In this work we hence perform first--principles simulation of $Mn$ $\delta$-doped $GaAs$ structures~\cite{AuthorGaAsMn_Exp1}, including the effect of the SO in a non perturbative way and using fully--relativistic PP. We then compute the absorption coefficients of the system and the MOKE parameters in the so-called polar geometry, where the magnetization vector is oriented perpendicular to the reflective surface and parallel to the plane of incidence. This is the most studied geometry and the one which gives the largest Kerr rotation.

	\begin{figure}[t]
	 \centering 
	 \includegraphics[scale=0.18]{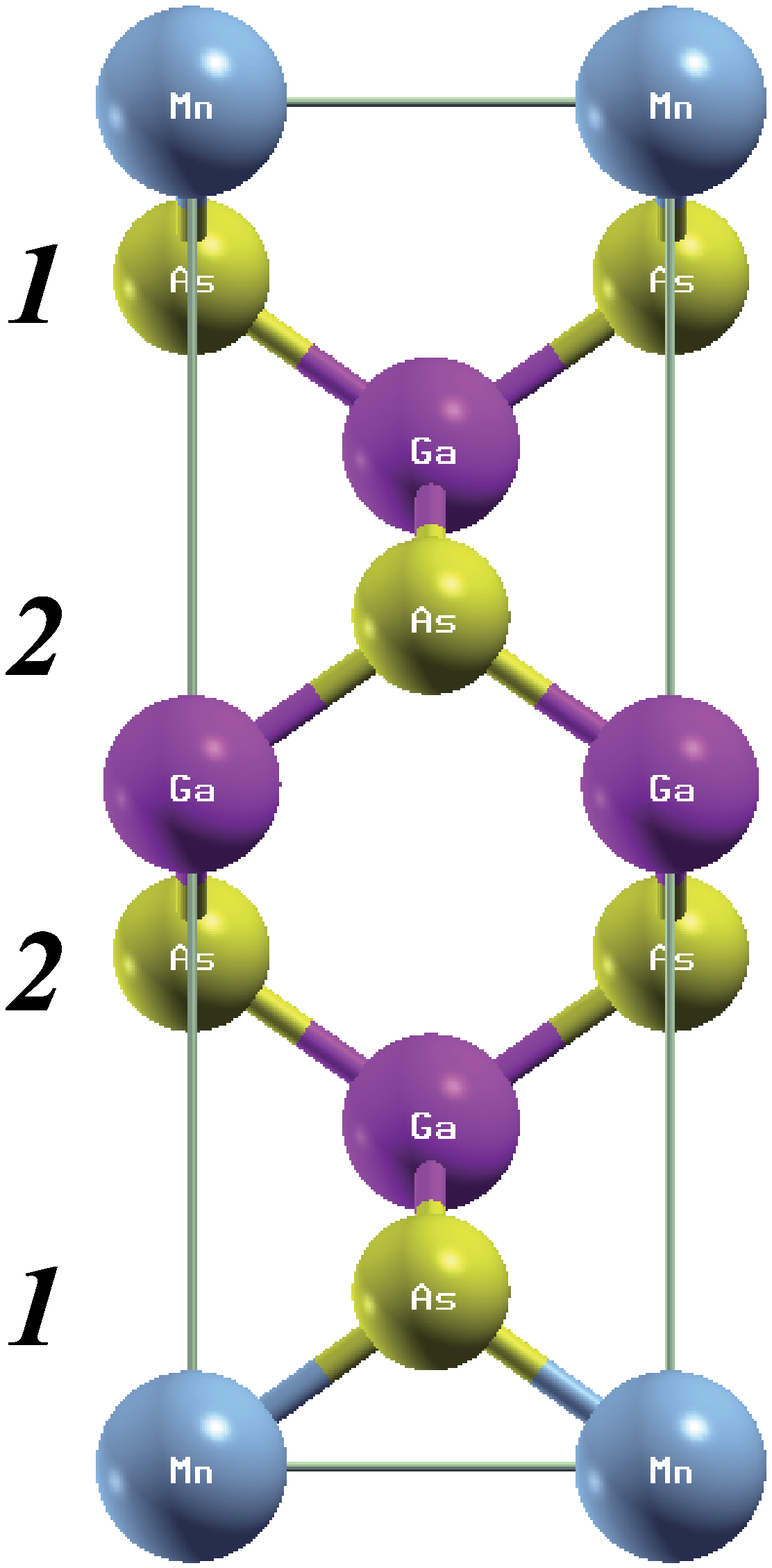}
	 \includegraphics[scale=0.28]{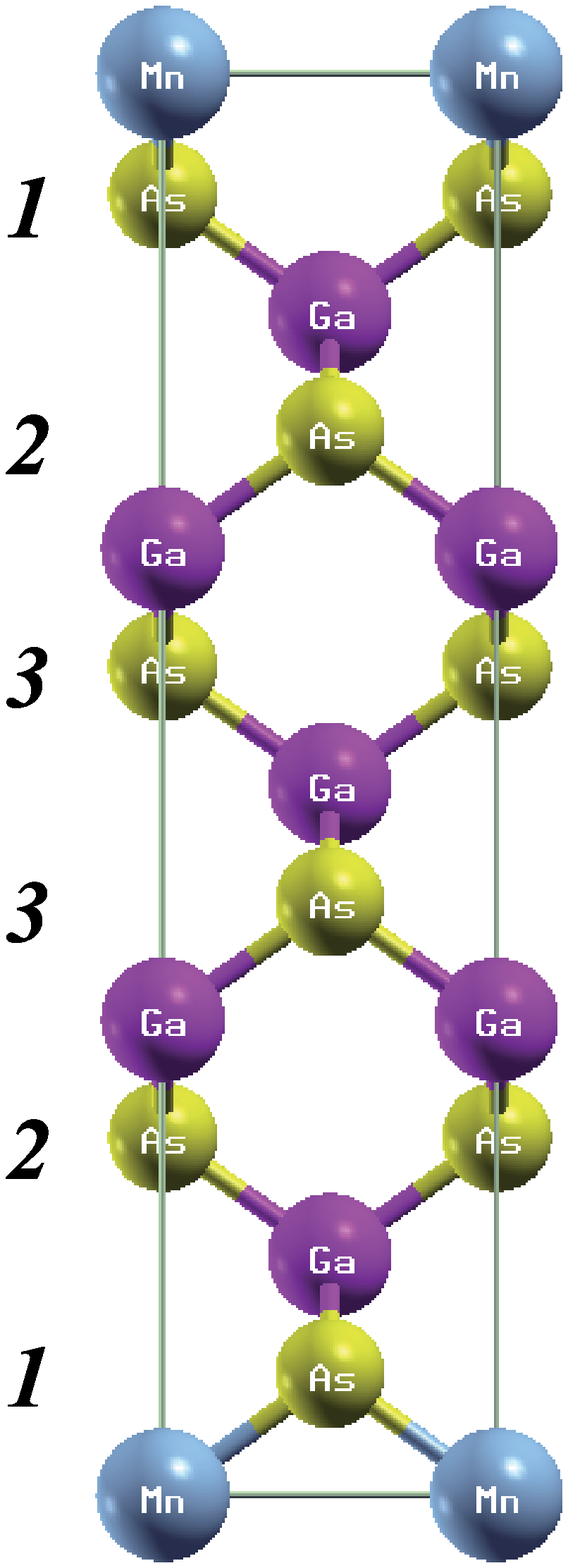}
	 \includegraphics[scale=0.28]{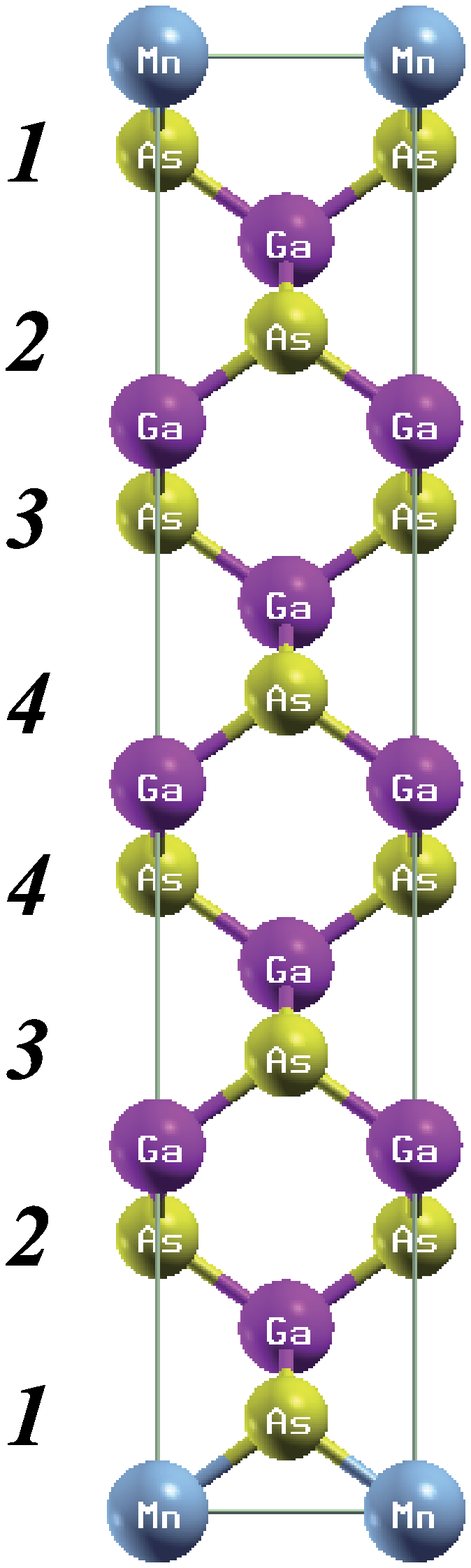}
	 \caption{(color online) Atomic structure (unit cells) of some of the studied $\delta$-doped materials. Yellow spheres are As atoms, the violet ones represent  Ga, and the cyan color is used for Mn. The corresponding Mn-doping concentrations are: $12.5\%$ (rightmost structure, an 8-layers super--cell with 7 GaAs layers separating MnAs ones);  $16.7\%$ (6-layers super--cell with 5 GaAs layers); $25\%$ (leftmost, 4-layers super--cell with 3 GaAs layers);  $50\%$ (2-layers super--cell, not shown). 
	 }
	 \label{fig:mat}
	\end{figure}

	This work is organized as follows. In Sec. \ref{sec:first}  we present DFT--LDA ground state results and Kohn--Sham (KS) band structure for heterostructures described by 8,6,4, and 2-layers super--cells (SC), corresponding to doping concentrations from $12.5\%$  to $50\%$
	(see Fig.~\ref{fig:mat}).
	We show the electronic density of states, projected onto spin components, atomic species, and $s$/$p$/$d$ angular momenta. In section~\ref{sec:spet} we deal with optical absorption spectra, computed within the linear response scheme in the random--phase approximation (RPA), and compared with experiments. We then extend our analysis to magneto--optical effects in section \ref{sec:kerr}, after briefly summarizing the basic equations for the MOKE and MCD spectra, and we discuss our results analyzing the similarities and differences with the case of uniform doping.  Finally, in section~\ref{sec:conclusions} we draw our conclusions.

	\section{Electronic properties of $\text{Mn}$ doped $\text{GaAs}$: delta doping} \label{sec:first}

	\begin{figure}[t]
	 \centering
	 \includegraphics[scale=0.31]{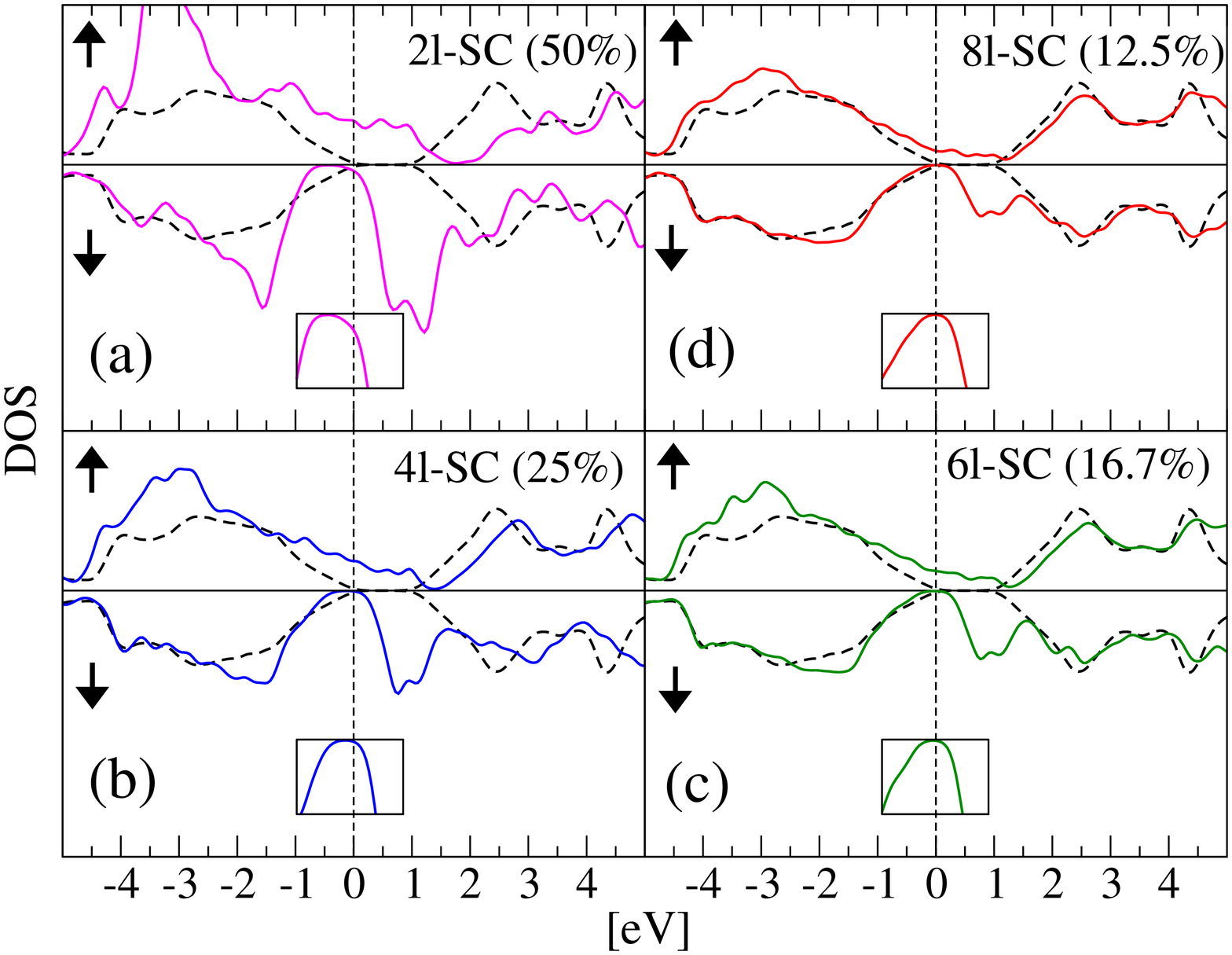}
	 \caption{(color online) Spin--resolved  Density of states  of the four 
	  studied $Mn$--doped $GaAs$ heterostructures, compared with that  of bulk 
	  GaAs (black dashed line). Panel $a$ to $d$ correspond to decreasing 
	  Mn concentrations, from $50 \%$ to $12.5\%$. 
	  Spin up and spin down components are computed by projecting the spinorial wavefunctions onto the S$_z$ spin eigenstates. 
	  The inset of each figure shows a zoom in the vertical scale  
	  of the density of states 
	  in the minority spin channel close to the Fermi level.}
	 \label{fig:2dos}
	\end{figure}

	We model $Mn$ $\delta$-doped $GaAs$ by means of the periodic super--structures
	shown in Fig.~\ref{fig:mat}. The variable thickness of bare $GaAs$ interposed between
	$MnAs$ layers allows us to simulate $\delta$-doped materials with 12\%, 16\%,
	25\% and 50\% of $Mn$ atoms replacing $Ga$ ones, (with the latter case 
	corresponding to an hypothetical $GaAs$/$MnAs$ super--lattice). 

	We perform our calculations with DFT-LDA \cite{AuthorDFT1,AuthorDFT2,AuthorDFT3} in the plane waves PP scheme, as implemented in the Quantum Espresso (QE) package~\cite{QuantumEspresso}.   
	To keep into account the spin--orbit effects on the valence wave-functions, the use of fully relativistic (norm conserving) PP is necessary.
	Moreover, the presence of non--collinear spin polarization (i.e., spinorial KS wavefunctions) must be allowed. Both features are available in QE.
	We use tetragonal unit cells, by relaxing the lattice parameter along the growth direction (perpendicular to the $Mn$ doped layers) and all the atomic positions within the unit cell~\cite{Footnote1}. The in--plane lattice parameter is kept fixed at its bulk $GaAs$ value, consistently with an epitaxial growth on the $GaAs$ (001) oriented substrate.
	Ground state calculations are performed with a kinetic energy cut---off of 50 Ry (680 $eV$) for the wave--functions,  and
	a Monkhorst--Pack~\cite{MP} grid 12x12xn for the Brillouin zone (BZ) with $n=8,4,3,2$ for the two, four, six and eight layers SC respectively,
	thus maintaining an (approximately) uniform mesh in the BZ. The ground state of bulk $GaAs$, which is also considered for reference, is computed using a 12x12x12 sampling.

	\begin{figure}[t]
	 \centering  
	 \includegraphics[scale=0.31]{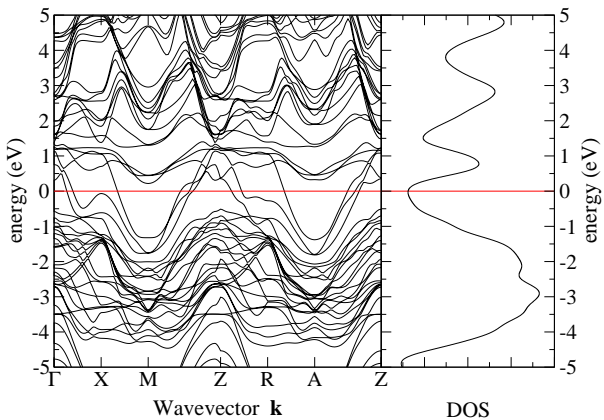}
	 \caption{(color online) Kohn--Sham band structure and total DOS for the 
	  25$\%$ doping case (panel $(b)$ of figure \ref{fig:2dos}). The red line is 
	  drawn at the Fermi energy.}
	 \label{fig:4band}
	\end{figure}

\begin{figure}[t]
	 \centering
	  \includegraphics[scale=0.3]{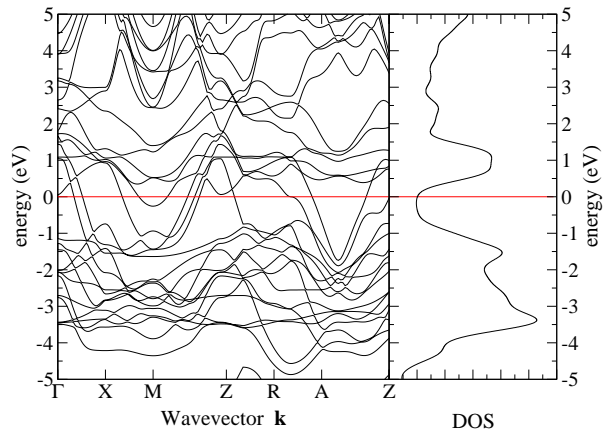}
	 \caption{(color online) Kohn--Sham band structure and total DOS for the 
	  50$\%$ doping case (panel $(a)$ of figure \ref{fig:2dos}). The red line is 
	 drawn at the Fermi energy.}
	  \label{fig:2band}
	  \end{figure}

	We compare  the density of states at the KS level~\cite{dos_details}, for the four systems considered, in figure \ref{fig:2dos}.
	All the $Mn$--containing systems have a metallic character, as expected. The case of undoped bulk $GaAs$ is also shown for comparison. Doped systems clearly display a nonzero DOS at the Fermi level, with a characteristic $Mn$--related structure lying in the top region of the bulk bandgap. The 50\% doped structure (panel $(a)$) clearly deviates from the behaviors shown by lower--doping systems, both at the level of the DOS and in term of Fermi surface topology. This difference can be clearly seen in the band structure, where the number of bands appearing to cross the Fermi level increases from three (see e.g. Fig.~\ref{fig:4band}), for the lower doping cases, to four (see Fig.~\ref{fig:2band}) in the  50\% doped system. At the same time, the DOS of the 50\% doped case displays a clear reduction of the spectral weight around 2.5 $eV$ above E${_F}$, typical of bulk $GaAs$, in favor of lower energy structures.

	Moreover, the total magnetic moment per unit cell is an integer multiple of $\mu_B$ (namely, $4\mu_{B}$) in all the structures except the 50\% doped one, where it drops slightly below the value of 4 $\mu_B$. A non-integer value of  $\mu$ / $\mu_B$ implies that the system cannot be an half metal, as also appearing from spin projected DOS (Fig.~\ref{fig:2dos}), where both the spin-up and spin-down component of the corresponding DOS are non-negligible at the Fermi energy.   
	In the inset of each panel we display the  
	spin-down component of the DOS,  zoomed in the vertical scale.  
	
	The 12\%, 17\%, and 25\% DFHs present a gap in the spin-down projected DOS, which is zero at the Fermi energy.  At variance, the spin-up DOS of all these 
	DFHs is clearly nonzero at the Fermi energy.
	Hence, the  electron population at the Fermi energy in the 12\%, 17\%, 
	and 25\% DFH is fully spin-up polarized, and these systems are half-metals, similarly to the case of uniformly $Mn$ doped $GaAs$ compounds  \cite{ReviewDMS-Sato1}.

	In view of the qualitative difference of the 50\% doped system, in the following discussion and in optical properties calculations only materials with integer  $\mu$ / $\mu_B$ will be considered, and compared with the uniformly doped systems.

	 \begin{figure}[t]
	  \centering
	  \includegraphics[scale=0.31]{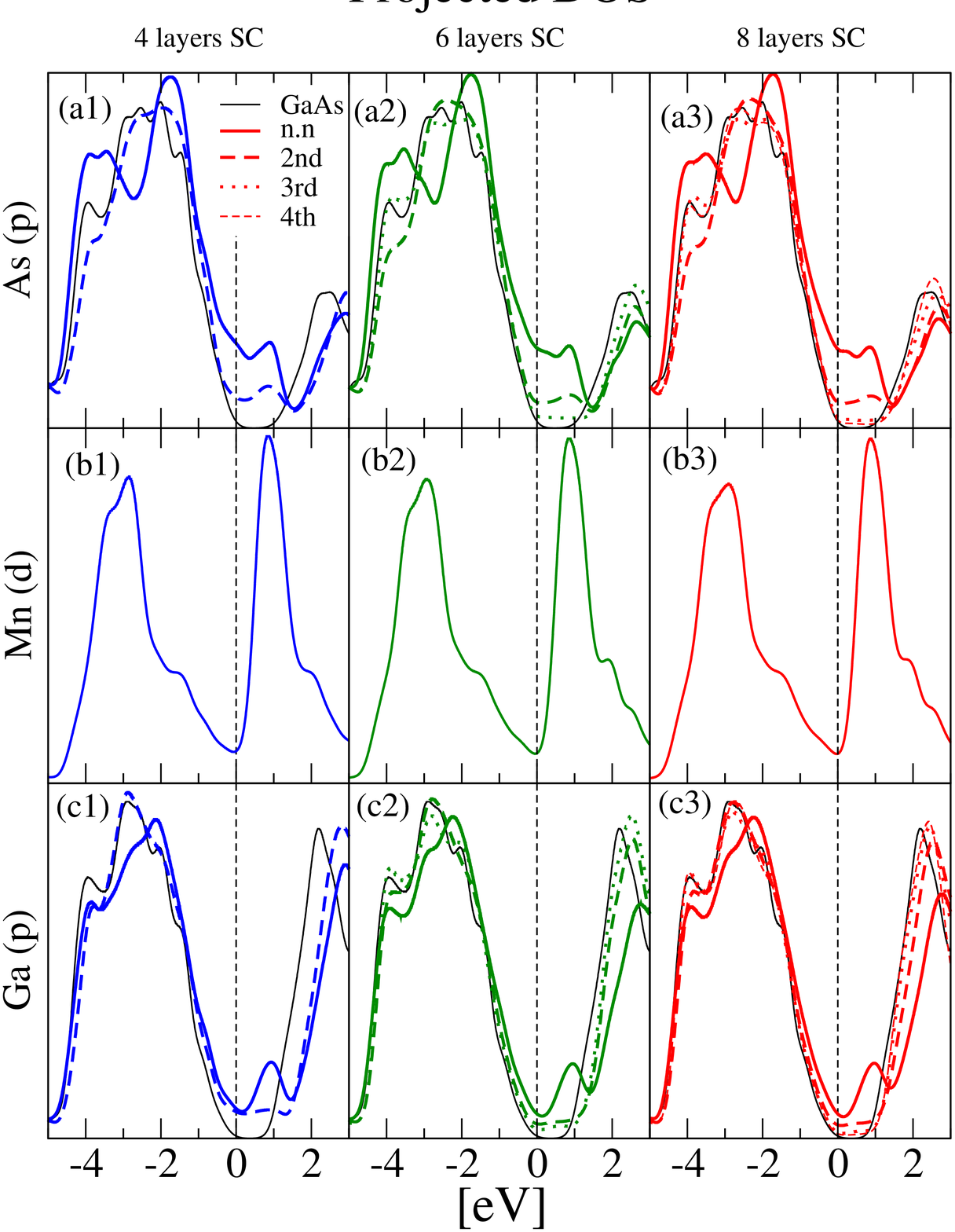}
	   \caption{(color online) Projected density of states of the studied MnGaAs heterostructures. The projections on $As$ $p$ orbitals (panels $a1$-$a3$), $Mn$ $d$ orbitals (panels $b1$-$b3$) and $Ga$ $p$ orbitals (panels $c1$-$c3$) are shown for the structures with 3 (blue), 5 (green) and 7 (red) $GaAs$ layers between the $Mn$ layers, respectively.
	   The projection on $As$ and $Ga$ species is atom resolved. Atoms are selected according to their distance from the $Mn$ layer, as
	   1st nearest neighbor (thick continuous line), 2nd (thick dashed line), 3rd (dotted line) and 4th (thin dashed line). 
	The case of bulk $GaAs$ is also shown for comparison (thin continuous black line).}
	   \label{fig:pdos}
	  \end{figure}

\begin{figure}[t]
	\centering
	\includegraphics[scale=0.5]{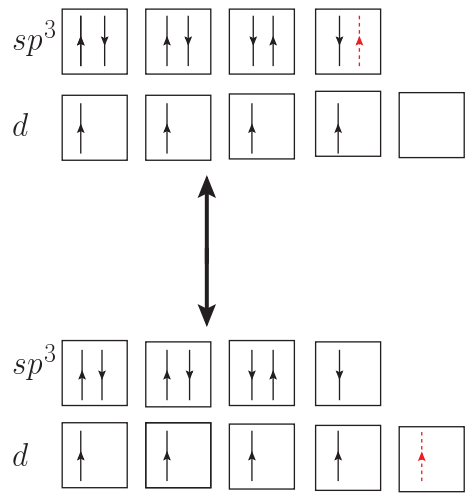}
	\caption{Schematic electronic diagram for the $As$--$Mn$ binding represented as two resonant configurations. Each $Mn$ orbital has four first nearest neighbor $As$ atoms. In first approximation, an hole is shared between the four $Mn$--$As$ $sp^3$ bonds orbitals and the $Mn$ $d$ orbitals. In practice the hole delocalize also on the other $sp^3$ orbitals of the doped $GaAs$ and is responsible of the ferro--magnetic interaction between different $Mn$ layers.}
	\label{fig:spin}
	\end{figure}

	A more detailed analysis of the band character can be performed by decomposing the DOS via a projection on the different atomic species and angular momenta. The major contribution to the metallic bands comes from $Mn$ $d$ orbitals and $As$ and $Ga$ $p$ orbitals. Fig. \ref{fig:pdos} shows the $As$ $p$ (top panel), $Mn$ $d$ (middle panel), and Ga $p$ (bottom panel) contribution for the three systems considered (i.e. three, five and seven layers super--cell). The $As$ and $Ga$ $p$ contribution to the undoped $GaAs$ DOS is also shown for comparison (black line).  The $p$ $As$ and $Ga$ contributions are also decomposed according to the distance of the considered atom from the $Mn$ layer, with the thick continuous line corresponding to the $As$ ($Ga$) nearest to $Mn$. As shown in figure \ref{fig:pdos}, the presence of $Mn$ mainly affects the first two closest layers of $As$ (labeled with $1$ and $2$  in Fig.~\ref{fig:mat}, thick continuous and dashed lines in Fig.~\ref{fig:pdos}) and the first closest layer of $Ga$ (same conventions). They all display a peak above the Fermi energy at the same energy of the $d$ orbital of $Mn$. These peaks indicate an hybridization of $d$ orbitals of $Mn$ with $p$ orbitals of the $As$ and $Ga$ atoms, or better with the $sp^3$ backbone of the $GaAs$ lattice. Indeed the $Mn$ atoms, replacing $Ga$ atoms, are ``forced'' to create an $sp^3$ hybridization. However, while $Ga$ has three electrons in the $s$ plus $p$ orbitals, $Mn$ only has two. There is then a competition between filling the hole in the $sp^3$ backbone and keeping five valence $Mn$ electrons in its $d$ orbitals. This competition is represented as two resonant configurations in Fig.~\ref{fig:spin}. The dashed red arrow indicates the electron shared between $sp^3$ and $d$ orbitals. In both configuration the magnetic moment is $4\mu_{B}/cell$. 

	Remaining  $As$ and $Ga$ atoms show a PDOS contribution very similar to the one in bulk $GaAs$, confirming that they are not substantially influenced by the presence of the $Mn$ dopant. This result confirms that the ferro--magnetic interaction between the $Mn$ layers becomes much weaker if they are separated by more than 3 $GaAs$ layers. However also at greater distance a small projection of the state immediately above the Fermi level remains on the $As$ and $Ga$ orbitals, suggesting that a smaller ferro--magnetic interaction is still present also in the 12\% doped system, where the $Mn$ layers are separated by 7 $GaAs$ layers.
	  
	\section{Optical properties of $\text{Mn}$ doped $\text{GaAs}$} \label{sec:spet}
	Optical absorption is proportional to the imaginary part of the macroscopic dielectric function, $\epsilon_{M} \l(\omega\r)$.
	The latter can be obtained from microscopic quantities such as the response function $\chi_{G,G'} (\bf{q},\omega)$ \cite{Onida1,Strinati}.
	Response functions can in turn be computed on the basis of transition matrix elements between electronic states (both interband and intraband in the case of metals). Several levels of theory can be adopted, ranging from the simplest approach, i.e. the bare independent particle--random phase approximation (IP-RPA), to more complete approaches where, e.g., local fields and/or excitonic effects are taken into account.
	In the long wavelength limit, which is appropriate for UV-VIS spectra, $\epsilon_{M}$ can be expressed as:

	\beq
	\label {eq:epsM}
	\epsilon_{M} \l(\omega\r)= 1 - \lim_{\bf{q}\rightarrow 0} \frac{4 \pi e^2}{|\bf{q}|^2} \chi (\bf{q},\omega).
	\enq

	Involving only the $G=0$, $G'=0$  ``head'' of the {\em appropriate} response function
	${\chi_{G,G'} (\bf{q},\omega)}$ written in reciprocal space \cite{Onida1}.  

	In this work, we use ${\chi^{RPA}_{\bf{G}\bf{G}'}(\bf{q},\omega)}$, i.e. the response function
	constructed including the local--field (LF) effects:

	\beq
	\label{rispchi}
	\begin{split}
	\chi^{RPA}_{\bf{G}\bf{G}'}(\bf{q},\omega)=&\chi^{QP}_{\bf{G}\bf{G}'}(\bf{q},\omega)+ \\
	&+\sum_{\bf{G}_1\bf{G}_2}\chi^{QP}_{\bf{G}\bf{G}_1}(\bf{q},\omega)
	f^{H}_{\bf{G}_1\bf{G}_2}\l(\bf{q},\omega\r)\chi^{RPA}_{\bf{G}_2\bf{G}'}(\bf{q},\omega)
	\end{split}
	\enq
	Here ${f^{H}_{\bf{G}_1\bf{G}_2}\l(\bf{q} \omega\r) = \delta v_H[\rho]/\delta\rho}$, the functional derivative of the Hartree potential, is the bare Coulomb interaction and corresponds to an electron--hole exchange term \cite{Onida1}.

	Neglecting ${f^{H}}$ one has ${\chi_{\bf{G}\bf{G}'}(\bf{q},\omega) =  [\chi_0^{QP}]_{\bf{G}\bf{G}'}(\bf{q},\omega)}$,
	i.e., the response function in the IP--RPA approximation.

	The quasi--particle (QP) response function an be expressed in terms of single--particle electronic eigenstates
	$\ket{\psi^{QP}_{n\bf{k}}}$ and eigen--energies ${\epsilon^{QP}_{n\bf{k}}}$ as:
	  
	\beq
	\begin{split}
	\chi^{QP}_{\bf{G}\bf{G}'}(\bf{q},&\omega)= -\frac{1}{V} \sum_{n\bf{k}}\sum_{m\bf{k}'}\l( f_{n\bf{k}} -f_{m\bf{k}'}\r) \\
	& \l( \frac{\bra{ \psi^{QP}_{m\bf{k}'}}e^{i \l(\bf{q}+\bf{G}\r) \bf{r}}\ket{\psi^{QP}_{n\bf{k}}} 
	 \bra{\psi^{QP}_{n\bf{k}}}e^{-i \l(\bf{q}+\bf{G}'\r)\bf{r}'}\ket{\psi^{QP}_{m\bf{k}'}}}
	{\epsilon^{QP}_{m\bf{k}'} -\epsilon^{QP}_{n\bf{k}}-\omega-i\eta} \r).
	\end{split}
	\enq

	${\epsilon^{QP}}$ are obtained from KS eigenvalues applying a gap-opening/ band-stretching correction
	which simulates the self--energy effects in semiconductors/metals, and $\ket{\psi^{QP}_{n\bf{k}}}$ are taken as unperturbed
	KS eigenstates $\ket{\psi^{KS}_{n\bf{k}}}$, as implemented in the YAMBO\cite{Yambo} code.

	\begin{figure}[t]
	 \centering
	 \includegraphics[scale=0.3]{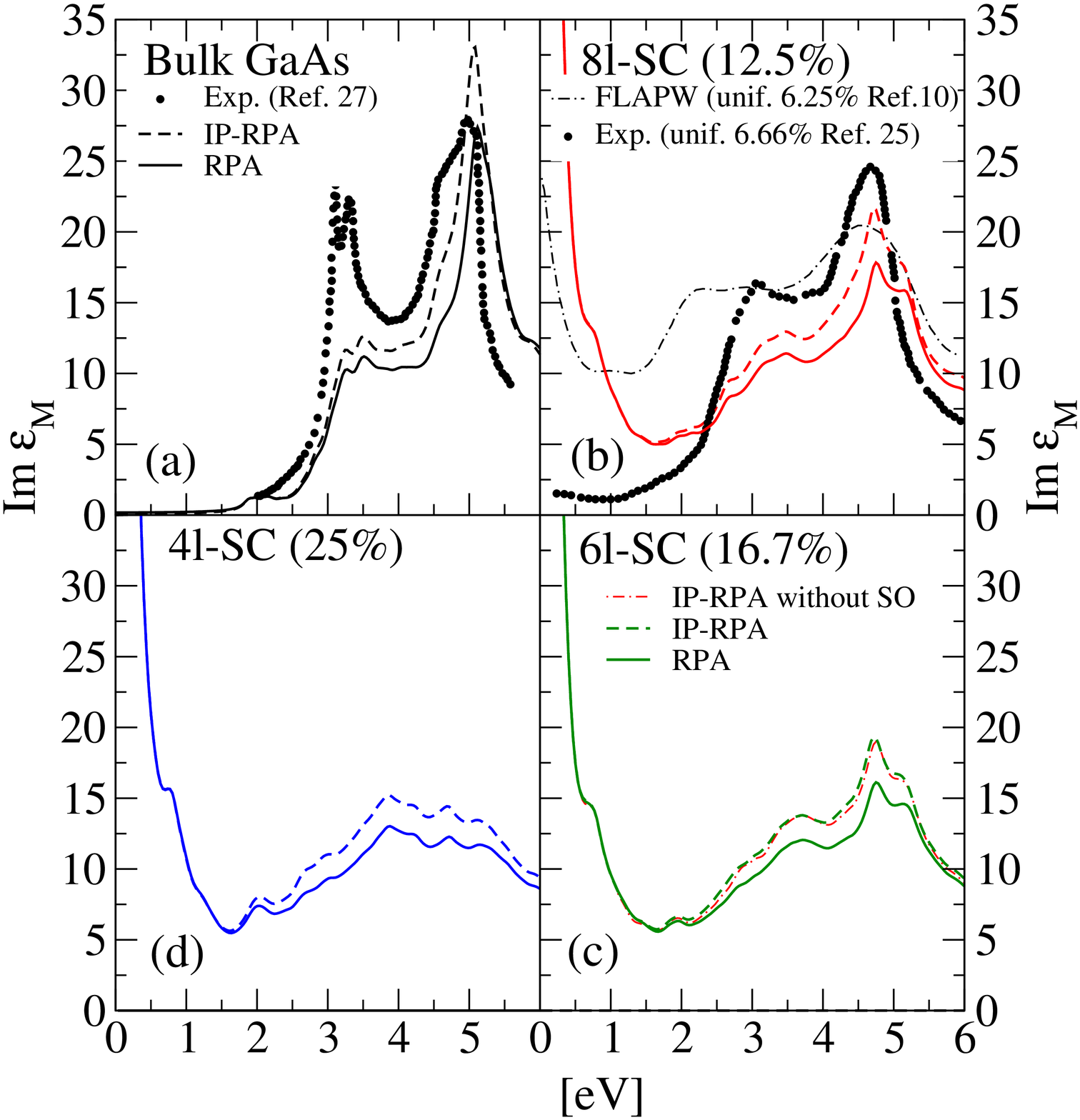}
	 \caption{(color online) Computed UV-VIS absorption spectra for light polarization parallel to the $Mn$ layers. Panel $(a)$ shows the bulk $GaAs$ case as reference, and $Mn$ $\delta$-doped heterostructures 
	with $12.5\%$, $16.7\%$, and $25\%$ Mn concentration are shown in panels $(b)$, $(c)$, and $(d)$, respectively. 
	The dashed lines and continuous lines represent the IP-RPA and  RPA results, respectively. Theoretical spectra are compared with experimental data for bulk $GaAs$ (circles in panel $(a)$, from Ref.~[\onlinecite{AuthorBulkGaAs_Exp1}]) and for a uniformly
	Mn-doped GaAs sample at low doping concentrations (dots in panel $(b)$, from  
	Ref.~[\onlinecite{Singley1}]). We also compare with the FLAPW ab--initio results of Ref.~[\onlinecite{Stroppa1}], computed for a uniformly doped system with a concentration of $6.25 \%$ (dot--dashed line).
}
 \label{fig:spec}
\end{figure}

Here we report the optical properties of our four, six and eight layers SC
of $Mn$ doped $GaAs$. The absorption spectra of the undoped system is also shown for reference as for the case of the DOS.
To obtain converged spectra a k--points sampling of the BZ with $24\times24\times n$ mesh is used, with $n=8,6,4$ for the four, the six layers and the eight layers respectively, considering $nv$ valence bands and $nc$ conduction bands in total. The Drude term is modeled with a plasma frequency $\omega_{p}=2.75 eV$ and relaxation frequency of $0.7 eV$ and the QP corrections are included as a $10\%$ stretching of the band structure, as in Ref.~[\onlinecite{Stroppa1}]. LF effects are found to converge including  $\bf{G}$ vectors in Eq.~\ref{rispchi} up to a kinetic energy a cutoff of $4 Ry$.  The absorption of bulk $GaAs$ is instead computed with a double-grid sampling of the BZ~\cite{AuthorDoubleGrid} with a regular $12\times12\times12$ mesh shifted from $\Gamma$ plus a random mesh of $2000$ k--points. In this case the well established value~\cite{AuthorQPcorr} of ${0.8\ eV}$ opening of the band gap has been used for the QP correction.

\begin{figure}[t]
\centering
\includegraphics[scale=0.3]{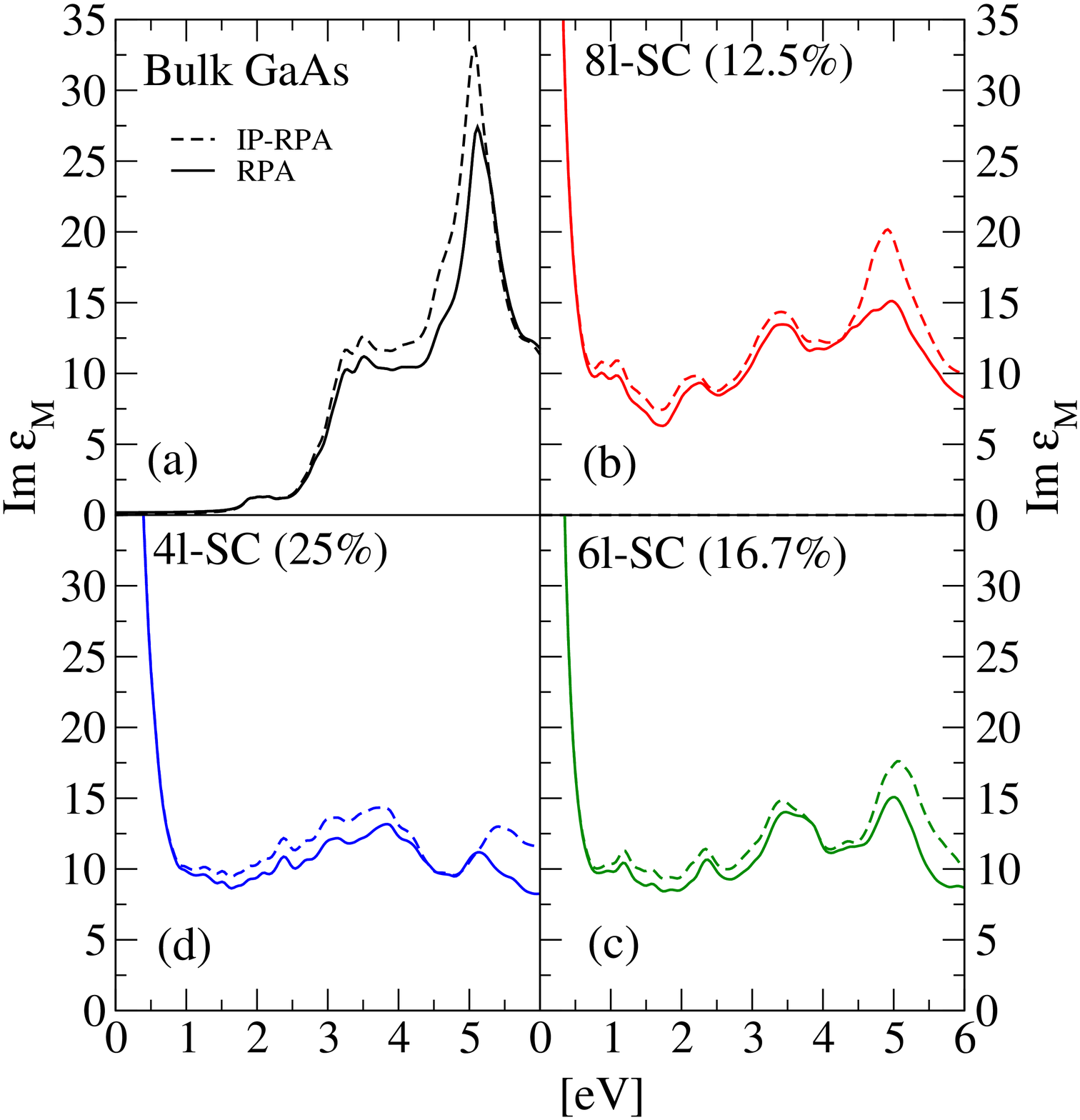}
 \caption{(color online) 
Computed UV-VIS absorption spectra of bulk $GaAs$ (panel $(a)$) and $Mn$ $\delta$-doped heterostructures with $12.5\%$ (panel $(b)$),
$16.7\%$ (panel $(c)$), and $25\%$ (panel $(d)$) Mn doping, for light polarization perpendicular to the $Mn$ layers (grazing incidence).
As in figure \ref{fig:spec}, the dashed lines and continuous lines represent IP-RPA and  RPA results, respectively.}
\label{fig:rot}
\end{figure}

In Figs.~\ref{fig:spec}-\ref{fig:rot} the absorption spectra, ${Im[\epsilon_{M}(\omega)]}$,
for the 12, 17 and 25\% doped materials is shown. The case of bulk (undoped) $GaAs$ is in panel $(a)$ of both figures.
Dashed lines show the IP--RPA spectra for comparison. 
The RPA spectra including LF effects differ from IP-RPA ones because the intensity of the absorption is reduced. The intensity reduction is peak-specific, and is due to the well known depolarization effect, which is usually more important in non uniform systems. However, LF do not change substantially the position of the peaks.

We first consider incident light with polarization on the layers plane (Fig.~\ref{fig:spec}).
The intensity of the main E2 peak around 5 $eV$ is lower
in DFHs than in bulk GaAs, and it decreases by increasing the Mn concentration.
In fact this peak becomes a double peak by effect of the Mn $\delta$-doping (panel $(b)$), and more structures appear by decreasing  the distance between Mn layers (panels $(b)$, $(c)$ and $(d)$). 
We compare in panel $(b)$ our IP-RPA results for the 12\% delta-doped systems (seven layers of Ga between the $Mn$ layers) with experimental data (circles in panel $(b)$) for an uniformly doped $Ga_{1-x}Mn_xAs$  sample with $x=0.066$ (Ref.~[\onlinecite{Singley1}]).   
The theoretical underestimation of the intensity of the 5 $eV$ (E2) peak is likely due to the different doping concentration between theory and experiment. Indeed the intensity of the E2 peak decreases increasing the dopant concentration, both experimentally (Refs.~[\onlinecite{Singley1,Burch1}]) and theoretically (present work, and Ref.~[\onlinecite{Stroppa1}]). 
We notice that the double peak structure at $\sim$ 5 $eV$ of Fig.~\ref{fig:spec} is peculiar of the $\delta$-doped $GaAs$ absorption spectra in polar geometry, since it is absent in the computed spectra with light-polarization perpendicular to Mn layer (Fig.~\ref{fig:rot}), as well as in the experimental data for randomly distributed Mn doping. The double peak also appears when the absorption spectra is computed without SO as shown in Fig.\ref{fig:spec}.c.2.

The influence of local field effects in the case of grazing incidence (Fig.~\ref{fig:rot}) is more pronounced for energies above 5 $eV$, where LF appear to suppress the peak intensity more than at lower energies.

On the other hand,
in both Figs.~\ref{fig:spec} and~\ref{fig:rot} the intensity of the 
theoretical  $\sim$ 3 $eV$ is underestimated, due to our neglect 
of excitonic effects\cite{AuthorQPcorr,AuthorBulkGaAs_Exp1}. 
The latter, however, are expected to be strongly suppressed in systems with efficient free-carrier screening, such as in metals or even half-metals, with respect to the case of a pure semiconductor.  In fact the peak at 3 $eV$ in the experimental data for uniform Mn doping (dots in panel $(b)$ of Fig.~\ref{fig:spec} ) is clearly less prominent than in bulk GaAs. 

The overestimation of  $\epsilon_M$ at very low energy in our simulation can be attributed, at least in part, to an overestimation of intraband transitions, described by the Drude term. The  discrepancies between experimental data and our results can also be partially accounted for by the different type of doping (uniform vs $\delta$-doping).

Moreover experiments are performed at diluted Mn concentration (few percents or less).  
When the concentration of $Mn$ in uniformly doped samples is sufficiently low, the overlap of impurity wave-functions is  
negligible and the compound is semiconducting.  At variance, in $\delta$-doped samples, the conductivity in the $Mn$ layers
remains metallic also at overall low doping concentrations.

\subsection{Kerr effect and reflection magneto--optical dichroism} \label{sec:kerr}

MOKE experiments evaluate the difference in  optical reflectivity for left and 
right circularly polarized light.
In practice, since a linearly polarized light can be thought as the result of the
superposition of left and right circularly polarized waves having the same 
intensity and phase, a nonzero difference in the dielectric response of the material  
can be detected as a change in the polarization properties of the reflected light. 

In this way, Kerr parameters are defined in terms of  the Kerr rotation ${\theta_K}$
and the Kerr ellipticity ${\eta_K}$, the first describing the angle between 
the major axis of the ellipse with respect to the original linear polarization
direction, and the second one being related to the ratio between the two ellipse axis.

In the present manuscript we consider the MOKE in the so called polar geometry, where the light propagates
along the direction of the magnetization of the sample, which we identify here as $z$,
perpendicular to the sample surface. The polarization hence lays in the $xy$ plane.
We now introduce the complex refractive index $n_{\pm}=\sqrt{\epsilon_{\pm}}$,
where  $\epsilon_+$
($\epsilon_-$) is the dielectric function for the right (left) circularly polarized light.
Then following Ref.~\onlinecite{kojima} 
the Kerr parameters are defined as: 
\beq \label{eq:kerr_param}
\theta_K\l(\omega\r)+i\eta_K\l(\omega\r)=
i \frac{ n_{+}(\omega)-n_{-}(\omega) } { n_{+}(\omega) n_{-}(\omega) -1 }.
\enq

${\epsilon_{\pm}}$ can be derived, for a cubic system, from the $xx$ and $xy$ elements of the dielectric
tensor, according to the expression: ${\epsilon_{\pm}=\epsilon_{xx} \pm i\epsilon_{xy}}$. 
The magneto--optical response can hence be computed starting from the 
diagonal and off-diagonal components of the dielectric tensor.
Indeed, if $\epsilon_{xy} \ll \epsilon_{xx}$, which is the common case, Eq.\ref{eq:kerr_param}
is well approximated by:
\beq \label{eq:kerr_param_small}
 \theta_K\l(\omega\r)+i\eta_K\l(\omega\r)\simeq
 \frac{-\epsilon_{xy}\l(\omega\r)}{\l[\epsilon_{xx}\l(\omega\r)-1 \r]\sqrt{\epsilon_{xx}\l(\omega\r)}}.
\enq
which holds for small Kerr angles. 

Remarkably, in such differential measurements, the 
contribution stemming from the off--diagonal elements of the dielectric tensor,
which is usually negligible in bare absorption or reflectivity experiments, becomes
important. $\epsilon_{xy}$ can be obtained from the response function
$\chi\left(\bf{q}_{\alpha}\bf{q}_{\beta} \omega \right)$, by generalizing Eq. \ref{eq:epsM}.
In practice, we evaluate the dielectric tensor by means of the YAMBO~\cite{Yambo} code, using the same
approach described in Ref.~[\onlinecite{Sangalli1}], where $\epsilon_{\alpha\beta}$, is computed, at the RPA--IP level, as 

\beq
\epsilon_{M\alpha \beta}\left(\omega\right)=\bf{1}-\lim_{\underset{\bf{q}_{\alpha} \rightarrow 0}{\overset{ \bf{q}_{\beta} \rightarrow 0}{}}} \frac{4 \pi e^2}{q^2}\chi\left(\bf{q}_{\alpha}\bf{q}_{\beta} \omega \right).
\enq 

${\epsilon_{xy}}$ must be converged against the same parameters needed to converge the absorption.
Convergence in the sampling of the BZ requires a
$24 \times 24 \times n$ grid where the number of k-points along the $z$ direction are $n=12,8,6$
for the 4, 6 and 8 layers super--cell respectively i.e. a slightly finer sampling along the
$z$ direction, if compared to the one used for absorption. The reason stems from the fact  that
${\epsilon_{xy}}$ is two orders of magnitude smaller than ${\epsilon_{xx}}$.
The knowledge of ${\epsilon_{\alpha\beta}}$ also allows one to compute
the reflectance ${R_{\pm}(\omega)}$ at normal incidence, defined as the square modulus of the
complex reflectivity, both
for right ($+$) and left ($-$) circularly polarized light~\cite{kojima}:
\beq
\label{eq:Rpm}
 R_{\pm} = \left \|   \frac{n_{\pm}(\omega)-1 }{ n_{\pm}(\omega) +1 } \right \|  ^2 .
\enq

Following Ref.~\onlinecite{kojima}, the latter quantities determine 
the reflectance magnetic circular dichroism (R-MCD) spectrum as:
\beq
R_{MCD}(\omega) = \frac{1}{2} \frac{ R_{+}-R_{-} }{ R_{+} + R_{-}},
\enq
which is more easily measured experimentally than the MOKE parameters.
In the limit of small Kerr angles one can prove, after some 
tedious but straightfoward algebra, that  
 $R_{MCD}(\omega)$ and Kerr ellipticity are brought to coincide: 
\beq
R_{MCD}(\omega) \approx  \eta_K(\omega),
\enq

\begin{figure}[t]
 \centering
 \includegraphics[scale=0.3]{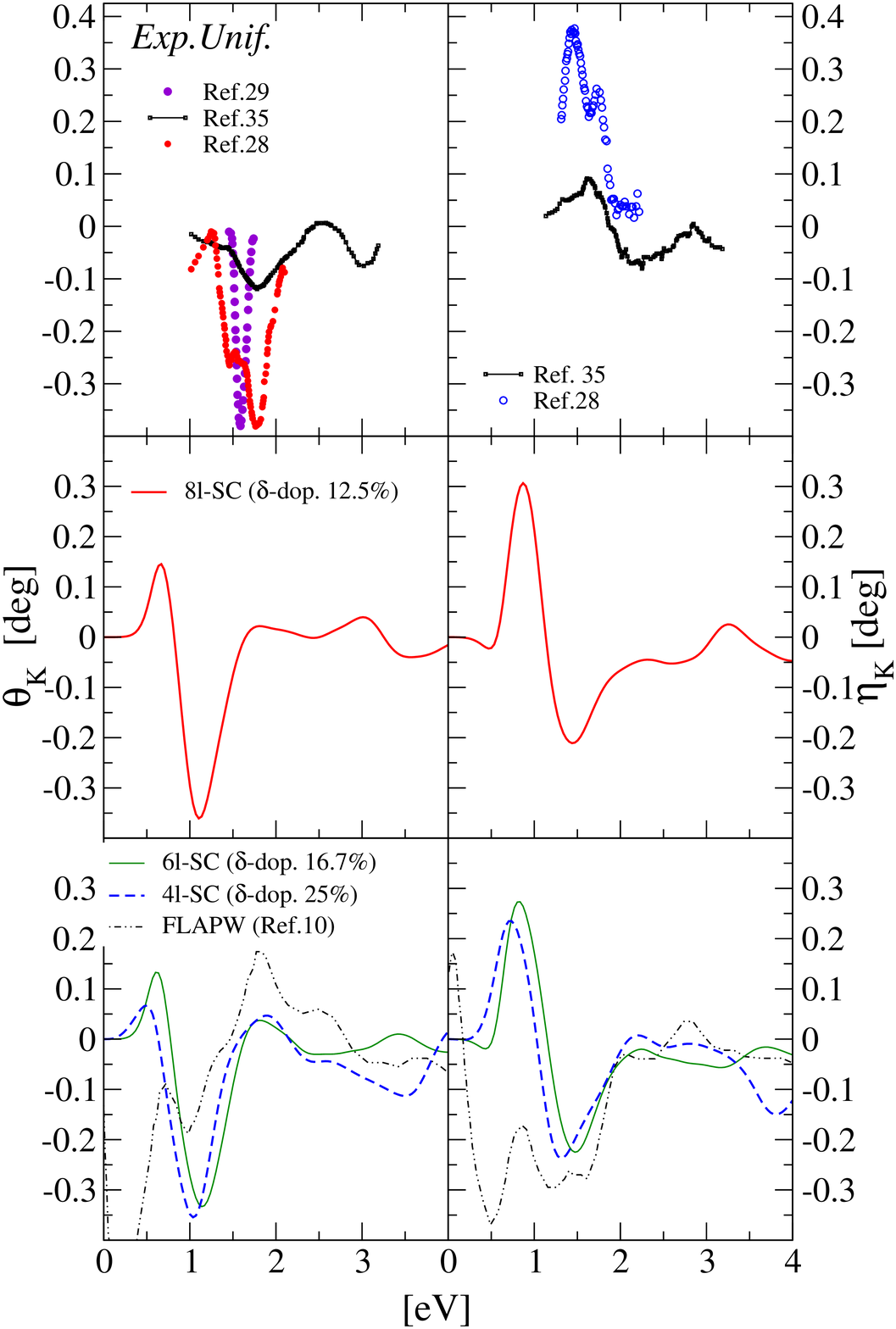}
 \caption{Kerr parameters computed at the IP-RPA level for three $Mn$ 
 $\delta$-doped $GaAs$ heterostructures. Central panels show the spectrum
 of the 12.5$\%$ Mn DFH (red line); the results for the $16.7\%$ and 
 $25\%$ Mn concentrations are shown in bottom panels (green and dashed blue 
 lines, respectively). 
 In the top panel, we show, for comparison, 
 several experimental data for uniformely doped samples: 
 black dots are for a $6\%$ Mn sample considered in 
 Ref.~[\onlinecite{ Kojima}], the red dots are for a $3\%$ Mn 
 case (Ref.~[\onlinecite{Furdyna1}]), and violet dots refer to 
 a $2\%$ Mn sample  (Ref.~[\onlinecite{Kimel1}])
 Finally blue empty dots in right top panel are experimental data for the Reflection
 MCD in a uniformely doped sample with $3\%$ $Mn$ from Ref.~[\onlinecite{Furdyna1}].
 In bottom panels, we report as dashed lines the results of FLAPW (LDA+stretched bands) calculations 
 for an uniformly doped systems, from Ref.~[\onlinecite{Stroppa1}] 
}
 \label{fig:Kerr_parameters}
 \end{figure}

In Fig.~\ref{fig:Kerr_parameters} we plot the computed 
Kerr parameters ${\theta_K}$ and ${\eta_K}$ 
for the considered DFH. We also show, by comparison, some 
theoretical and experimental literature data for uniformely doped MnGaAs.

We first observe that there is no clear dependence of the computed Kerr parameters on the Mn doping concentration. This finding is in agreement with previous calculations for the uniform doping case\cite{Picozzi1,Stroppa1}. Our calculations shows that,
as it happens for the absorption, the computed MOKE is similar to the one obtained for the uniformly doped case.
In Fig.~\ref{fig:Kerr_parameters} (bottom panels)
the Kerr rotation is compared with all--electrons calculations for the uniform system at the LDA level~\cite{Stroppa1}.
Despite this difference in the calculations, different structures (uniform doping and DFH) and different doping concentrations all calculations and experimental data shows at least one 
common dominant feature, that is a negative Kerr rotation at low energy. In the experimental data
for uniform doping~\cite{Furdyna1,Kimel1} this negative peak in the Kerr rotation is blue--shifted by $\approx 0.6\ eV$ with respect to our results
(see Fig.~\ref{fig:Kerr_parameters}. top left panel). This shift also appears in the computed Kerr ellipticity/R-MCD data (top right panel) and in DFT calculations for the uniform doping case. It maybe due to inadequacy of the $10\%$ band stretching approach to mimic QP corrections, in particular for the Mn-$d$ empty states. The latter, being only $1\ eV$ above the Fermi level, are essentially unaffected by the stretching approach. 

We can thus conclude that at least the first peak in the Kerr rotation remains
substantially unchanged for doping concentrations varying from $12.5\%$ to $25\%$, and
it is barely affected by the details of the geometrical distribution of the $Mn$ atoms.

\begin{figure}[t]
 \centering
 \includegraphics[scale=0.3]{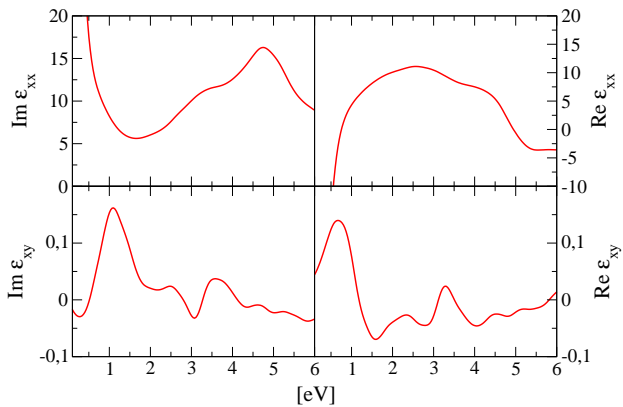}
 \caption{ Real and imaginary part of the diagonal and off-diagonal components of $\epsilon$ for the $Mn$ 
 $\delta$-doped $GaAs$ heterostructure corresponding to a 12.5$\%$ doping (Ref.~[\onlinecite{Furdyna1}])}
 \label{fig:eps_8loff}
 \end{figure}
 
As it appears from Eq.~\ref{eq:kerr_param_small}, a peak in the Kerr rotation can either have an ``optical'' origin, i.e.
it stems from a vanishing value of the denominator,
or a ``magneto-optical'' one, i.e. it arises from maximum/minimum of $\epsilon_{xy}(\omega)$ in the numerator. For the uniformly doped 
system the minimum close to $1\ eV$ has been reported to be of ``optical'' origin~\cite{Picozzi1,Stroppa1}.

To analyze if this is the case also for the DFH, we plot
in Fig.~\ref{fig:eps_8loff} the diagonal and the off--diagonal dielectric function.
We clearly see that $Im[\epsilon_{xy}]$ has a maximum at $~ 1\ eV$ and also that
both $Re[\epsilon_{xx}]$ and $Im[\epsilon_{xx}]$ are small in the same region.
So the corresponding peaks of the Kerr rotation, in the present case, have
both optical and magneto--optical origin.

Comparing the computed Kerr parameters with the computed DOS and 
its projections (see Figs.~\ref{fig:2dos}-\ref{fig:pdos}),
we see that a key role is likely played by the empty state induced by the $Mn$ doping,
as discussed in sec.~\ref{sec:first} (see also Fig.~\ref{fig:spin}).
Indeed we note that this state is located about 1 $eV$ above the Fermi level.
Thus the maximum (in absolute value) of the rotation is likely due to 
transitions from the valence band to this state, i.e. 
it is due to a transition involving both $d$ and $sp^3$ states. 

We thus suggest that in DFH the MOKE
could be substantially affected by electron/hole doping. Indeed, varying the  holes concentration, the
semi--metallic character of the system can be substantially affected, leading to a very different absorption
spectra, in particular in the region close or below $1\ eV$ as already suggested for the uniformly
doped system~\cite{Hankiewicz}.
This conclusion is also supported by previous all--electrons results for the uniform doping case,
where the Kerr spectrum is shown to change
significantly when $As$ anti--site defects are considered together with $Mn$ doping~\cite{Picozzi1}.

\subsection{Faraday effect and transmission magneto--optical dichroism} \label{sec:faraday}

\begin{figure}[t]
 \centering
 \includegraphics[scale=0.3]{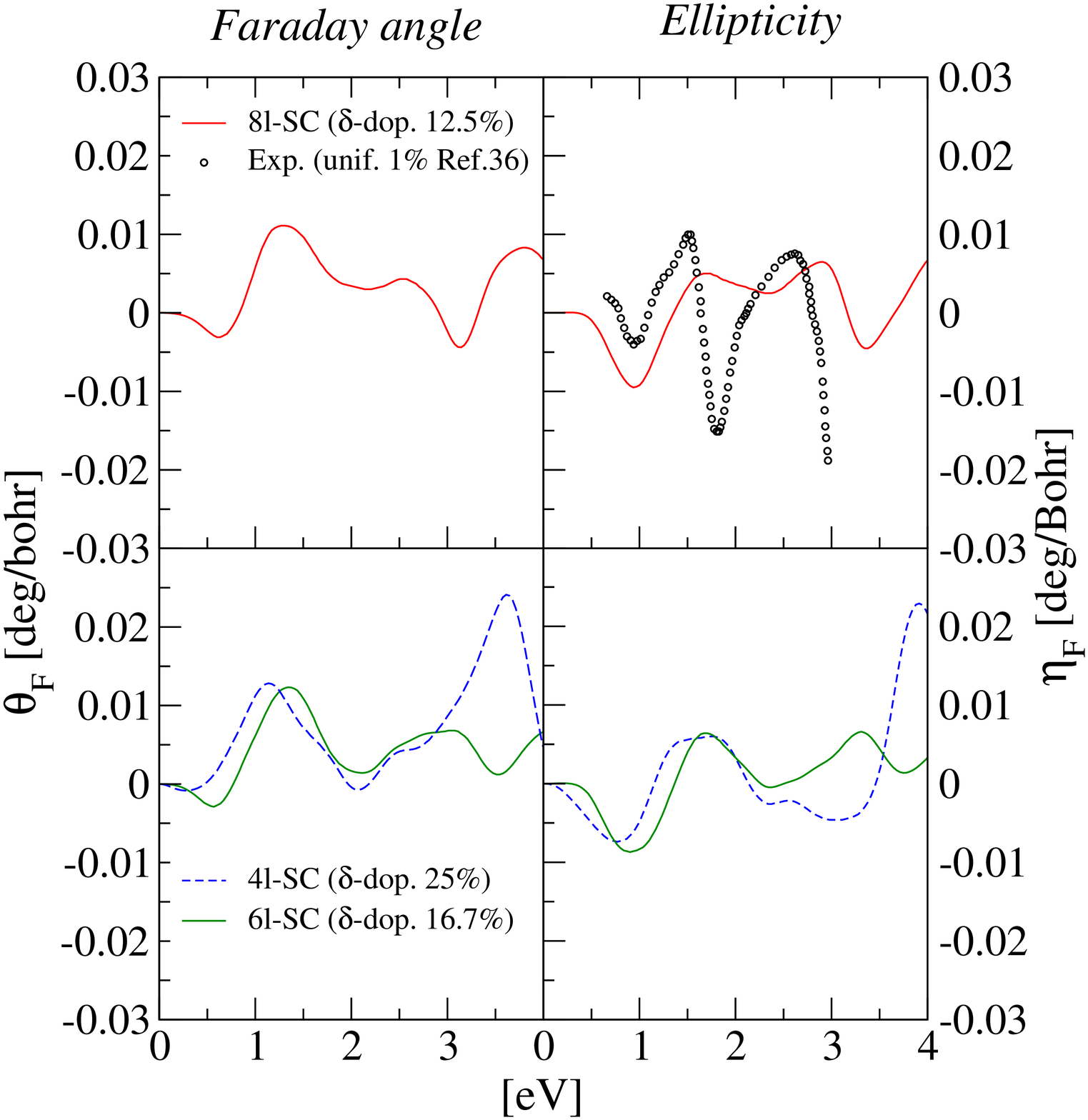}
 \caption{Faraday parameters / transmission MCD computed at the IP-RPA level for three $Mn$ 
 $\delta$-doped $GaAs$ heterostructures, corresponding to 12.5$\%$, 
 $16.7\%$, and $25\%$ Mn concentrations. Black empty dots are experimental data for uniformly 
doped samples with $1\%$ $Mn$, from Ref.~[\onlinecite{Ando2}].}
 \label{fig:Farad_parameters}
 \end{figure}

While the Kerr effect and Reflection MCD imply measuring the light 
reflected by a magnetic material, the corresponding effect for light 
transmitted through a magnetic medium is known as Faraday effect. 
In polar geometry (the one considered in the present study), 
when a linear polarized light propagates parallel to the magnetization of a ferromagnet, a Faraday rotation is observed as a rotation of the polarization plane of the transmitted light ($\theta_F$).  
In term of circularly polarized reflective indexes, the Faraday rotation $\theta_F$ and 
the Faraday ellipticity $\eta_F$ read\cite{kojima}:

\beq
\theta_F\l(\omega\r)+i\eta_F\l(\omega\r)=
i \frac{\omega}{2c}\left (  n_{+}(\omega)-n_{-}(\omega) \right ) L  
\enq
where $L$ is the sample thickness and $c$ the speed of light. 
We evaluated the Faraday parameters per unit length for the $Mn$ $\delta$--doped $GaAs$ in a similar 
way as done for Kerr parameters; our results are displayed 
in Fig.~\ref{fig:Farad_parameters}.
As the Kerr parameters are related to the R-MCD, the Faraday parameters are related to the
transmission MCD~\cite{Furdyna2} (T-MCD):
\beq
T_{MCD}(\omega) =
\frac{ T_{+}-T_{-} }{ T_{+} + T_{-}},
\enq
which, at normal incidence, is proportional to ${\Delta \alpha (\omega)}$, i.e.
the difference of the absorption coefficient ${\alpha_\pm(\omega)=2\omega Im[n_\pm(\omega)]/c}$ between  
the right and the left circularly polarized light: 
\begin{equation}
\Delta \alpha (\omega)= \alpha_+ (\omega)- \alpha_- (\omega)=
-\frac{4 \eta_F\l(\omega\r)}{L}
\end{equation}
Thus we can compare the results of our simulations with 
T-MCD experimental data for diluted concentration~\cite{Ando}
in randomly $Mn$--doped $GaAs$.

First thing that we observe is that, despite the physics of the Faraday effect / T-MCD is very similar to the on of the
Kerr effect / R-MCD, in this case the results are slightly more sensitive to the doping concentration. At least the region above
$2\ eV$ is affected going from $12.5\%$ to $25\%$ doping. However, as in the case of the Kerr, there is a structure slightly
above $1\ eV$, which is substantially unchanged varying the doping concentration.

To investigate possible applications of DFH as Faraday rotators,
in the following we compare some of our computed optical and 
magneto-optical data 
with the corresponding quantities of bulk transition metals,
which are typical textbook examples for the Faraday effect.\cite{coey}

We consider the 4-layers DFH and the peak at $\approx 1$ eV: 
according to our simulation, the Faraday rotation is 
, 
$\theta_F\simeq 60^\circ \mu m^{-1}$, 
significantly larger than the typical values for Fe, Co, and Ni, 
(35, 36, 10  $^\circ \mu m^{-1}$, 
respectively)\cite{coey}
(taking into account the $0.6\ eV$ red shift of our computed spectra
with respect to the experimental ones, these values are taken 
at $\lambda = 820\ nm$, i.e. at about $1.5\ eV$).
On the other hand, DFH have a lower optical absorption coefficient, which   
is desirable for efficient transmission. The optical penetration depth, 
defined as the length over which an electromagnetic wave of given frequency
encounters a 1/e reduction of its intensity --proportional to the 
inverse of the absorption coefficient--  is about 36 nm for  
our 4 layers DFH. For comparison, the optical penetration depth of 
transition metals are of the order of 15 nm.

\section{Conclusions} \label{sec:conclusions}

In conclusion, our results for Mn $\delta $-doped GaAs show that such digital ferromagnetic heterostructures have optical absorption and magneto-optical spectra displaying both similarities and differences with respect to previously studied uniformly doped MnGaAs systems.
The 50\% Mn doping case (one layer of GaAs between MnAs layers) corresponds to a system with qualitatively different properties, having lost the half-metallicity. 
All other structures, corresponding to 12-25\% Mn doping, have optical absorption spectra which change gradually with the $Mn$ concentration. The resulting spectra compare well with the available experimental data for the low--doping ($\leq 12\%$) system and theoretical data for the higher doping ($25\%$), for the uniformly doped case. 
With increasing Mn concentration the typical bulk GaAs spectral shape becomes less and less structured. In agreement with the experimental findings
the peak at $\simeq 3\ eV$ is strongly suppressed, likely because of the hole doping induced by the Mn impurities, which screens the electron--hole interaction responsible
for the creation of the excitonic peak. The peak at $\simeq 5\ eV$ is instead less affected.

Although the absorption spectra of the $\delta$-doped materials share the essential features of those of uniformly $Mn$ doped $GaAs$ 
at similar concentrations, the effect of local fields is enhanced, due to the stronger non-uniformity of the $\delta$-doped system. This effect is particularly evident when the light polarization has a large component along the growth direction, while it is less important
when the polarization lays in the plane of the $Mn$ layers.   
MOKE spectra do not show a clear trend with the doping concentration: essentially, the maximum value of Kerr rotation is constant, and quite similar to that of the uniform doped material~\cite{Stroppa1,Picozzi1}.
 Although the Faraday ellipticity is more sensitive to the doping concentration, also in this case the low energy spectrum remains substantially
unchanged moving from the high doping case ($25\%$) to lower doping.
On the other hand, the important role of the hole shared between the $Mn$ $d$ 
and $GaAs$ $sp^3$ states suggests that the studied DFH could be very 
sensitive to a further doping with electrons.
--------

\section*{Acknowledgments}
We acknowledge the CINECA and the Regione Lombardia award under the LISA initiative, for the availability of high performance computing resources and support, 
 the Cariplo Foundation for funding
(OSEA Project No. 2009-2552), 
and  R. Colnaghi for technical support. 
One of us (GO) acknowledges the ETSF-Italy\cite{matsuura_etsf2012} for support.  
 
\bibliographystyle{apsrev4-1}
\bibliography{paper}

\begin{thebibliography}{99}
\bibitem{Footnote1} The total residual force is below 1.7 mRy/Bohr
\bibitem{dos_details} The density of states has been computed projecting the wave--functions either on the $+S_z$ or the $-S_z$
spin channel. In this way we have been able extract the spin--resolved DOS from our calculations.
\end{thebibliography}

\end{document}